\documentclass[a4paper,aip,reprint,groupedaddress,showkeys,amsmath,amssymb]{revtex4-1}

\usepackage{dcolumn}
\usepackage{bm}
\usepackage{changebar,color,rotating,eso-pic,booktabs,fontenc}
\usepackage{graphicx}
\usepackage[mathlines]{lineno}

\definecolor{blue}{rgb}{0.2,0.2,0.4}
\definecolor{red}{rgb}{0.793,0.238,0.336}
\definecolor{white}{rgb}{1,1,1}
\definecolor{gray}{rgb}{0.85,0.9,1}

\begin{document}

\title{Long Range Node-Strut Analysis of Trabecular Bone Micro-architecture}
\author{Tanya Schmah}
\affiliation{
Department of Computer Science,
University of Toronto, \\
Toronto, Ontario M5S 3G4,
Canada
}
\author{Norbert Marwan}
\affiliation{Potsdam Institute for Climate Impact Research (PIK), 
D-14412 Potsdam, Germany, and\\
Interdisciplinary Center for Dynamics of Complex Systems, 
University of Potsdam, 
D-14415 Potsdam, Germany}

\author{Jesper Skovhus Thomsen}
\affiliation{Department of Connective Tissue Biology, Institute of Anatomy,
University of Aarhus, 
DK-8000, \r{A}rhus C, Denmark}

\author{Peter Saparin}
\affiliation{Perceptive Informatics, a PAREXEL technology company, 
Am Bahnhof Westend 15, 
D-14059 Berlin, Germany}


\begin{abstract}
{\bf Purpose}
We present a new morphometric measure of trabecular bone micro-architecture,
called \emph{mean node strength} (NdStr), which is part of a newly-developed
approach called
\emph{long range node-strut analysis}.
Our general aim is to describe and quantify the apparent 
``lattice-like'' micro-architecture of the trabecular bone network.

{\bf Methods}
Similar in some ways to the topological node-strut analysis introduced by
Garrahan et al.,
our method is distinguished by an emphasis on long-range
trabecular connectivity.
Thus, while the topological classification of a pixel (after skeletonisation)
as a node, strut, or terminus, can be
determined from the $3\times 3$ neighbourhood of that pixel,
our method, which does not involve skeletonisation,
takes into account a much larger neighbourhood.
In addition, rather than giving a discrete classification of each pixel
as a node, strut, or terminus,
our method produces a continuous variable, \emph{node strength}.
The node strength is averaged over a region of interest to produce the 
\emph{mean node strength} (NdStr) of the region.

{\bf Results}
We have applied our long range node-strut analysis to a set of 26 high-resolution
peripheral quantitative computed tomography (pQCT) axial images of human proximal
tibiae acquired 17~mm below the tibial plateau. 
We found that NdStr has a strong positive correlation
with volumetric trabecular bone mineral density (BMD). After an exponential transformation,
we obtain a Pearson's correlation coefficient of $r = 0.97$.
Qualitative comparison of images with similar BMD but with very different
NdStr values suggests that the latter measure has 
successfully quantified the prevalence of the ``lattice-like'' micro-architecture
apparent in the image.

Moreover, we found a strong correlation ($r = 0.62$), between NdStr and
the conventional node-terminus ratio (Nd/Tm) of Garrahan et al. 
The Nd/Tm ratios were computed using traditional histomorphometry performed
on bone biopsies obtained at the same location as the pQCT scans.

{\bf Conclusions}
The newly introduced morphometric measure 
allows a quantitative assessment of the long-range connectivity
of trabecular bone. One advantage of this method is that it is based on 
pQCT images that can be obtained noninvasively from patients, i.e.~without 
having to obtain a bone biopsy from the patient. 

\end{abstract}



\maketitle

\section{Introduction}

Several studies have shown that measures of trabecular
bone micro-architecture and bone strength are correlated.
\cite{goldstein1993,mosekilde1985,delling1995,rho1998,hildebrand1999,rajapakse2004,kinney2005}
Together with loss of bone mass, changes in the trabecular bone micro-architecture occur
during ageing, during development of osteopenia and osteoporosis as well as
in connection with immobilisation or space flight, and can lead to an increased risk of bone
fracture. 
The vertebral bodies and the epiphyses and metaphyses of the long bones consist mainly
of trabecular bone surrounded by a thin cortical shell. \cite{eastell1990,ritzel1997}
A dramatic change in the state of the trabecular bone leads to an increased fracture risk.
\cite{mcdonnell2007}

Bone mineral density (BMD) is the most commonly used 
predictor of bone strength and fracture risk, and also the most commonly used
general descriptor of the state of the bone.
Non-linear relationships have been established between volumetric BMD and
compressive bone strength and elastic modulus. \cite{carter1977,rajapakse2004}
However, for trabecular bone, it has been established that
a part of the variation in the strength of the bone cannot be explained
by BMD alone, but is instead due to the micro-architecture of the trabecular network.
For example, a relationship has been established between the mechanical properties
of the bone and the shape, orientation, bone trabecular volume fraction, and thickness 
of the trabeculae.
\cite{goldstein1993,mosekilde1985,ebbesen1999,jiang1998,majumdar1998,
odgaard1997,ulrich1999,baum2010,thomsen1998,thomsen2002a}
A series of new methodologies based on techniques from nonlinear data analysis
has also been introduced in order to study the relationship between
the complexity of the trabecular bone network and bone strength. \cite{saparin1998,dougherty2001b,prouteau2004,saparin2005, saparin2006,marwan2007pla,marwan2007epjst,marwan2009}
Saparin et al.~established such a relationship by use of
structural measures of complexity based on symbolic encoding. 
\cite{saparin1998,saparin2006}
Furthermore, several studies using numerical modelling of the trabecular bone and 
finite element analysis have also confirmed the importance of the
trabecular bone micro-architecture to bone strength. \cite
{rietbergen1998,guo2002,pothuaud2002,wolfram2009}

In the present study we propose a new method for analysis of trabecular bone
micro-architecture from high-resolution quantitative
computed tomography (QCT) images, as at present it is still not
possible to perform micro CT ($\mu$CT) on humans in vivo and in situ.
In contrast to classical
histomorphometry, CT images can be obtained in a nondestructive and
noninvasive manner, which is preferable
in a clinical setting.
Our method uses a new approach called
long range node-strut analysis that quantifies the apparent nodes and struts
in the trabecular network. 
In contrast to the node-strut analysis of Garrahan et al., \cite{Garrahan1986}
our method emphasises long range connectivity of the trabecular network,
over a distance controlled by a parameter in the algorithm.
The analysis has the dual aims of describing the shape of the trabecular network and predicting
bone strength.

The trabecular bone compartment consists basically of two components: bone and marrow.
Due to the limited resolution of present day CT and peripheral 
quantitative CT (pQCT) scanners it is not possible to completely resolve the trabecular
bone micro-architecture.
This results in variations of the CT values of the trabecular voxels, even if the intrinsic bone density is constant throughout the trabecular network. Consequently, our method takes this apparent variation in bone density into consideration rather than segment or binarise the image, which would imply a loss of this additional information. 
We assume that higher CT values of a given trabecula will, all else being equal, result in a higher compressive bone strength.
%
We note that our method does not require skeletonisation
of the image.

We apply our method to 2-dimensional pQCT images of human proximal tibiae 
and quantify the trabecular bone micro-architecture at different levels of bone integrity ranging
from normal healthy bone to osteoporotic bone (as assessed by their BMD).
We compare our results with  
the node/terminus ratio (Nd/Tm), \cite{Garrahan1986}
computed using traditional histomorphometry performed
on bone biopsies obtained at the same location as the pQCT scans.

\section{Materials} \label{sectMaterials}
The study population comprised 18 women aged 75--98 years and 8 men aged 57--88 years.
At autopsy, the tibial bone specimens were placed in formalin for fixation.
For each specimen a pQCT image and a bone biopsy were obtained from the same location. 

\begin{figure}[h] 
\centering \includegraphics[width=\columnwidth]{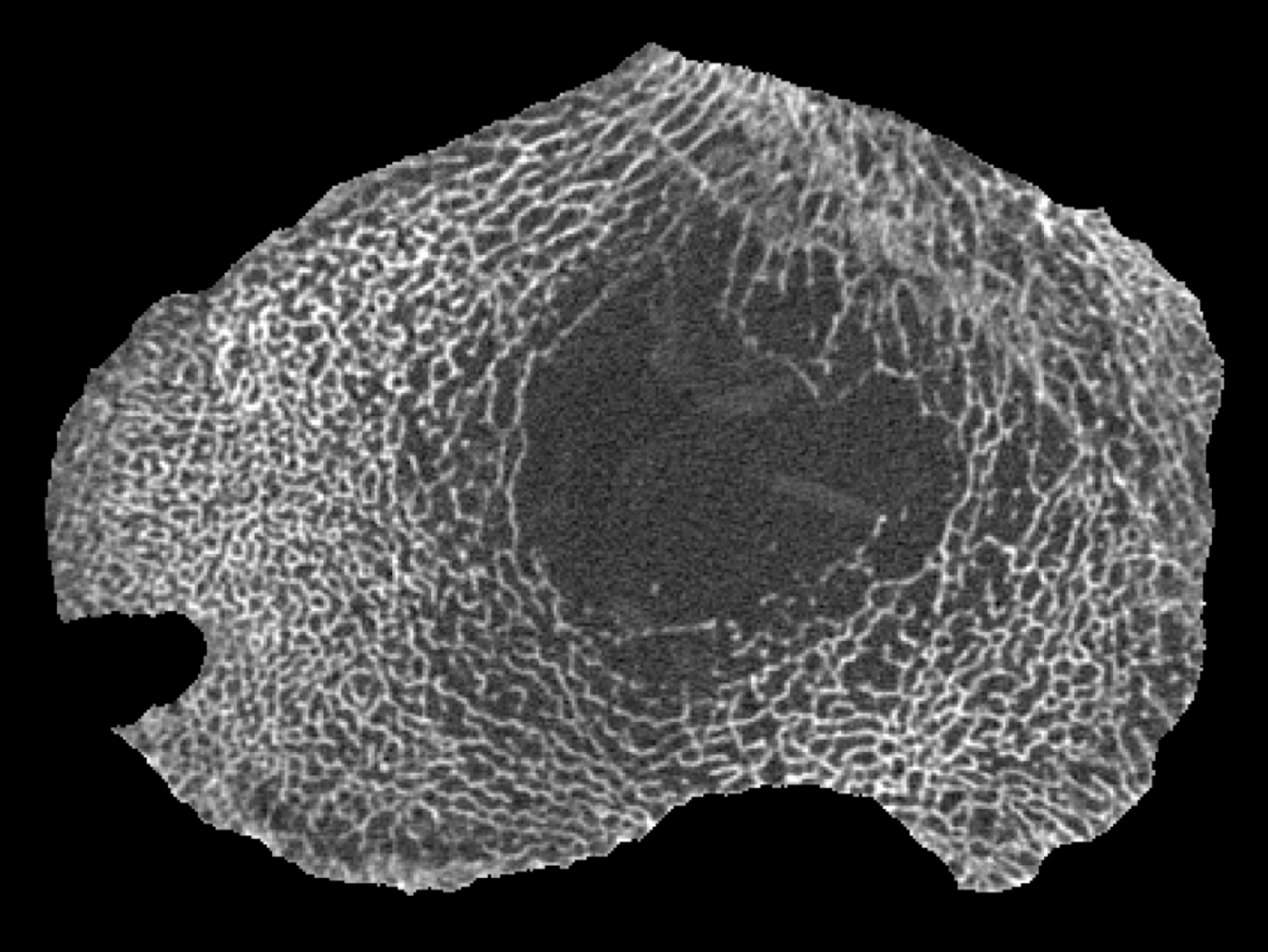}
\caption{
An axial pQCT slice of human proximal tibia acquired 17~mm 
below the tibial plateau. The cortical shell has been removed from the
image. The trabecular BMD of this sample is 107\,mg/cm$^3$.
The hole visible on the left of the image is the result of a cylindrical biopsy.}
\label{fig:tibia}
\end{figure}

For each proximal tibia, an axial QCT slice was acquired 17~mm below the
tibial plateau with a Stratec XCT-2000 pQCT scanner (Stratec GmbH, Pforzheim, Germany),
with an in-plane  pixel size of 200\,$\mu$m $\times$ 200\,$\mu$m and a slice
thickness of 1 mm.
In some cases, the scans were performed after the biopsies were taken.
Therefore, the holes
left from the bone biopsy appear in some of the pQCT images.
A standardised image pre-processing procedure was applied to exclude the 
cortical shell from the analysis. \cite{saparin1998,saparin2002,saparin2006}
One of the resulting images is shown in Fig.~\ref{fig:tibia}.
%

Cylindrical bone samples with a diameter of 7~mm were obtained 
17~mm distal from the centre of the medial facet of the superior articular
surface by drilling with a compressed-air-driven drill with a diamond-tipped
trephine at either the right or the left proximal tibia.
These bone biopsies were embedded undecalcified in methyl methacrylate, 
cut in 10-$\mu$m-thick sections on a Jung model K microtome
(R. Jung GmbH, Heidelberg, Germany), and stained with aniline blue
(modified Masson trichrome). The mounted sections
were placed in an flat-bed image scanner and 2540~dpi digital 1 bit images
of the sections were obtained as previously described in detail. \cite{thomsen2005}
The resulting  pixel size is 10 $\mu$m $\times$ 10 $\mu$m.



The trabecular BMD of each pQCT slice was calculated  
using a linear relationship derived
on the basis of
experimental calibration with the European Forearm Phantom,
as described by Saparin et al. \cite{saparin2006}
The trabecular BMDs of the slices in this study range from
30 to 150\,mg/cm$^3$, with a median of 97 \,mg/cm$^3$.

\section{Analytical Method} \label{sectAnalysis}

Most of this section describes the new image analysis method, 
\emph{long range node-strut analysis}, which includes the 
new measure, \emph{mean node strength} (NdStr).
This method was applied to the pQCT sections described in Section \ref{sectMaterials},
after removal of the cortical shell.
We also computed a standard measures from the same
regions of interest of the same pQCT images: 
trabecular volumetric bone mineral density (BMD), calculated as described 
by Saparin et al. \cite{saparin2006}

For further comparison, topological 2-dimensional node-strut analysis \cite{Garrahan1986} 
was performed on the histological sections described in Section \ref{sectMaterials},
using a custom-made computer program.
\cite{thomsen2000} The trabecular bone profile was iteratively eroded until
it was only one pixel thick using a Hilditch skeletonisation procedure, \cite{Lam1992}
and nodes and termini were automatically detected by inspecting the local $3\times3$
neighbourhood. 
If the centre pixel of the $3\times3$ neighbourhood is a skeleton pixel and one and only
one of the 8 other pixels is a skeleton pixel the centre pixel is classified as a
terminus (Tm) indicating that the strut ends in this pixel. If the centre pixel is
a skeleton pixel and three or more of the 8 adjacent pixels are skeleton pixels the
centre pixel is classified as a node (Nd) indicating that two or more struts join in
this pixel. Compston et al.~have argued that the ratio between nodes and termini
(Nd/Tm) is an expression of the connectivity of the trabecular network.\cite{Compston1995}
Consequently, only
the node-to-terminus ratio (Nd/Tm) from this analysis of the histological sections is used in the present study.

\subsection{Basic definitions}

We start with the algorithm to find \emph{strands} in each image.
A strand is a connected trabecular path, i.e., a chain of one or more struts, 
with the whole path going in approximately the same direction.
In our algorithm, we use eight directions, labelled by the points of the compass:
north (N), northeast (NE), east (E), southeast (SE), south (S), southwest (SW), west (W), or northwest (NW).
``North'' is the anterior direction, which corresponds to ``up'' in the images shown in this article.
A \emph{node} is a pixel that is joined to strands in at least three of these eight directions,
any two of which are at least 90 degrees apart (e.g. N, E, and SW;
but not N, NE, and E).
The \emph{node strength} of a pixel is 0 if it is not a node at all, but otherwise depends
on the lengths of the strands that meet at the node and the pQCT values of the pixels in these strands.

\subsection{CT values of bone}

These basic definitions could be implemented in a variety of ways,
depending for example on the definitions of ``connected'' and ``same direction'',
and on how the pQCT values of the pixels are used.
In our algorithm, 
the first step is to remove marrow from further consideration,
by applying a bone threshold filter.
The thresholded pQCT values will be referred to in this section as 
\emph{CT values of bone}, denoted $b$.
Thus if $a$ is the pQCT value of a pixel, then 
\begin{equation}\label{E:threshold}
b = 
\begin{cases}
a - a^{\textrm{threshold}} \quad \textrm{if} \quad a > a^{\textrm{threshold}}\\
0 \quad \textrm{otherwise}.
\end{cases}
\end{equation} 

We choose 
$a^{\textrm{threshold}} = 275$, corresponding to 
a BMD value of 24~mg/cm$^3$.
This is the soft tissue threshold used in symbol encoding
for complexity measures by Saparin et al. \cite{saparin2006} 
The threshold was determined as the pQCT value of the
densest non-bone tissue tested, plus the standard deviation of the noise (a calibration protocol for finding the threshold based 
on a phantom will be developed in the future). 


\subsection{Strands}

In order to explain the central algorithm, we begin by
considering a fictitious image consisting of just one row.
The CT values of the pixels in the row form a sequence
\[
b_0, b_1, b_2, b_3, \cdots  \,.
\]
We will define a corresponding sequence of \emph{strand strengths},
\[
s_0,s_1,s_2,s_3,\cdots \,,
\]
where each $s_j$ describes the pattern of densities to the left of pixel $j$ in the row.
We begin by setting $s_0 = 0$, and then 
we recursively define
\begin{align} \label{E:strandstr}
s_j = \left(b_{j-1} + T s_{j-1}\right) \left(\frac{\min\{b_j,b_{j-1}\}}{b_{j-1}}\right)\, , 
\end{align}
where $T$ is a \emph{transmission constant} between $0$ and $1$.
Note that if all of the CT values have a common value $b$ then
\begin{align*}
s_j &= \left(b + T s_{j-1}\right) 
=  \left(b + T \left(b + T s_{j-2}\right)\right) \\
&= b \left( 1 + T \right) + T^2  s_{j-2} = \cdots \\
&= b \left(1 + T +  \cdots + T^j \right), 
\end{align*}
since $s_0 = 0$. As $j$ increases,
$s_j$ approaches an upper bound of $\dfrac{b}{1-T}$.
Thus, for the simple case of a uniformly dense row, we have defined the
strand strength to be $\dfrac{1}{1-T}$ times the CT values of bone. 
This constant $\dfrac{1}{1-T}$
can be interpreted as a characteristic length of the method,
which we can vary depending on the length of strand that we believe to be
most important to the strength of the bone.
For the present study, we have chosen $T=0.95$, corresponding to a
characteristic length of $20$ pixels, i.e.~4~mm.
For the general case of a variable CT values row, the presence of
the transmission constant $T$ in the recursive formula ensures that 
the CT value of pixels closest to the $j^{\textrm{th}}$ pixel have the greatest
effect on $s_j$.
Finally, the factor $\dfrac{\min\{b_j,b_{j-1}\}}{b_{j-1}}$
ensures that $s_j$ depends strongly on $b_j$, so that any weak link
in a chain of high-density pixels lowers the strand strength dramatically.

\begin{figure}[h] 
\centering \includegraphics[width=\columnwidth]{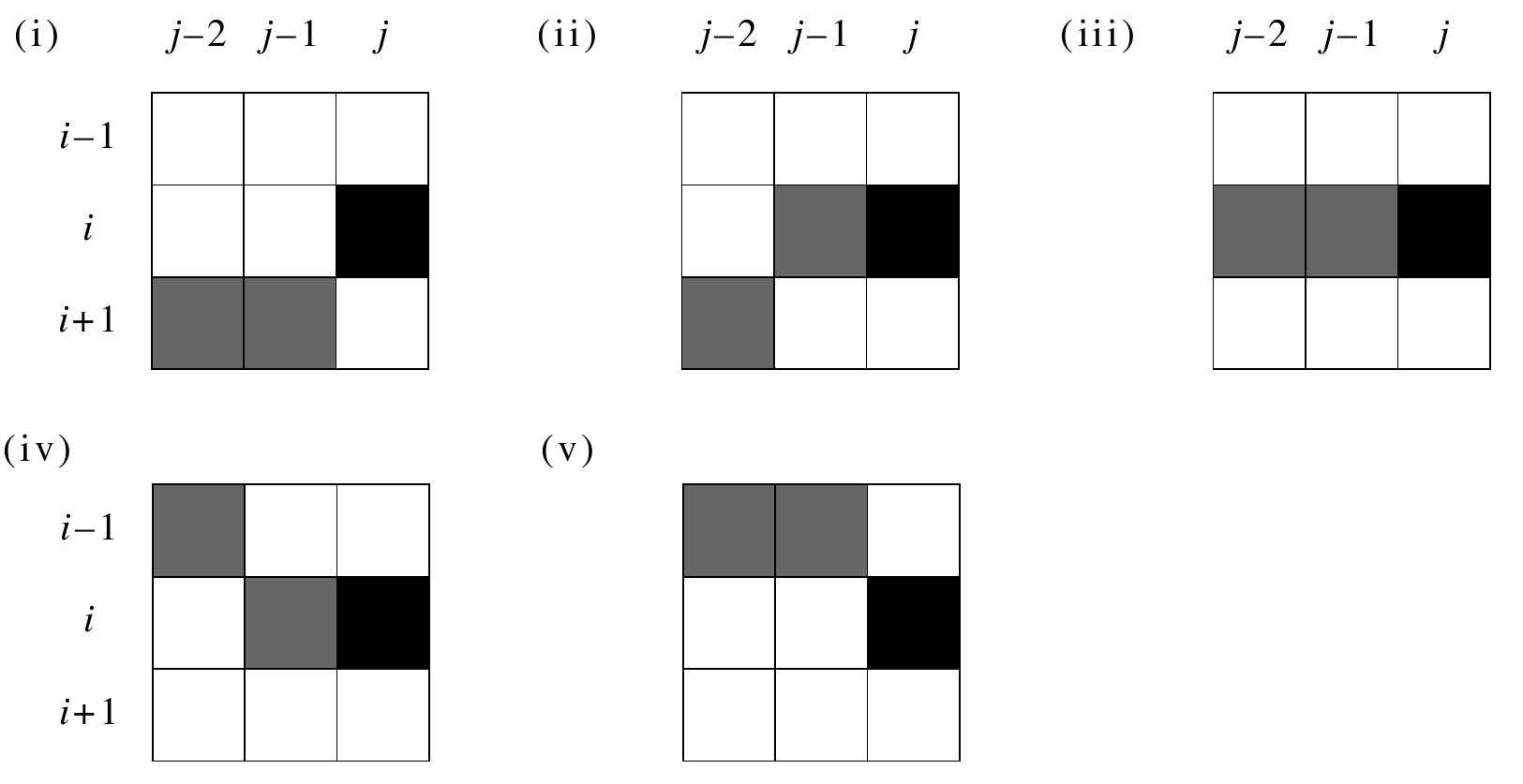} 
\caption{
The five possible length-3 paths in a leftwards (``west'') direction
from pixel $(i,j)$. The start pixel $(i,j)$ is marked in black.
}
\label{fig:path3}
\end{figure}

The difficulty in generalising this definition to a 2-dimensional image is
of course that there are infinitely many directions but only four that are 
parallel with the pixel grid.
We have resolved this in a practical but ad hoc fashion. 
As already stated, we consider eight directions. 
Consider first the definition of strand strength in the leftwards (W) direction.
>From pixel $(i,j)$ (at row $i$ and column $j$) 
we consider that there are five length-3 paths
leading approximately leftwards:
\begin{linenomath*} 
\begin{align*}
&(\text{i}) \hfill &(i+1, j-2)&\leftarrow (i+1,j-1) \leftarrow (i,j), \\
&(\text{ii}) \hfill &(i+1, j-2)&\leftarrow (i,j-1) \leftarrow (i,j), \\
&(\text{iii}) \hfill &(i, j-2)&\leftarrow (i,j-1) \leftarrow (i,j), \\
&(\text{iv}) \hfill &(i-1, j-2)&\leftarrow (i,j-1) \leftarrow (i,j), \\
&(\text{v}) \hfill &(i-1, j-2)&\leftarrow (i-1,j-1) \leftarrow (i,j), \\
\end{align*}
\end{linenomath*} 
as illustrated in Fig.~\ref{fig:path3}.


\subsection{Strand strength}

For each of these paths, we define 
$b_{j}, b_{j-1}$ and $b_{j-2}$ as
the CT values of the pixels in the path in columns 
$j, j-1$, and $j-2$. We then apply the recursive formula
\begin{linenomath*} 
\begin{align}\label{E:strandstr2}
s_j &=& \left(b_{j-1} + T \left(b_{j-2} + T s_{j-2}\right)
\left(\frac{\min\{b_{j-1},b_{j-2}\}}{b_{j-2}}\right)\right) \\ \nonumber
&& \times \left(\frac{\min\{b_j,b_{j-1}\}}{b_{j-1}}\right),
\end{align}
\end{linenomath*} 
which is the formula in Eq.~(\ref{E:strandstr}), applied twice. (We must
now set $s_1 = s_0 = 0$.) We now have five $s_j$ values, one for each of
the five paths, i.e., $s_j^{(i)},\ s_j^{(ii)},\ldots, s_j^{(v)}$. We
multiply all but the third (Fig.~\ref{fig:path3}(iii)) of these five
values by a \emph{bending coefficient} $\gamma$ between $0$ and $1$, to
penalise strands that bend away from the horizontal direction. Then the maximum of
these five numbers is our leftwards (W) strand strength $S^W_{ij} = \max
\{\gamma s_j^{(i)},\ \gamma s_j^{(ii)},\ s_j^{(iii)},\ \gamma s_j^{(iv)},\
\gamma s_j^{(v)}\}$.

Strand strengths in the other seven directions 
$S^{NW}_{ij},\ S^N_{ij},\ S^{NE}_{ij},\ldots, S^{SW}_{ij}$ 
are determined in a  similar manner. For
diagonal directions,  the geometry of the five length-3 paths used in the
computation is slightly different, so a different bending coefficient
$\gamma$ is used. We chose  $\gamma = 0.6$ for the horizontal and vertical
directions and $\gamma = 0.8$ for the diagonal directions. 
These constants were chosen so as to give mean strand strengths that are
approximately invariant to image rotations by an arbitrary angle, which were verified by
numerical experiments on tibial images.

\subsection{Node strength}

\begin{figure*}[htbp] 
\centering 
\includegraphics[height=.3\textwidth,angle=-90]{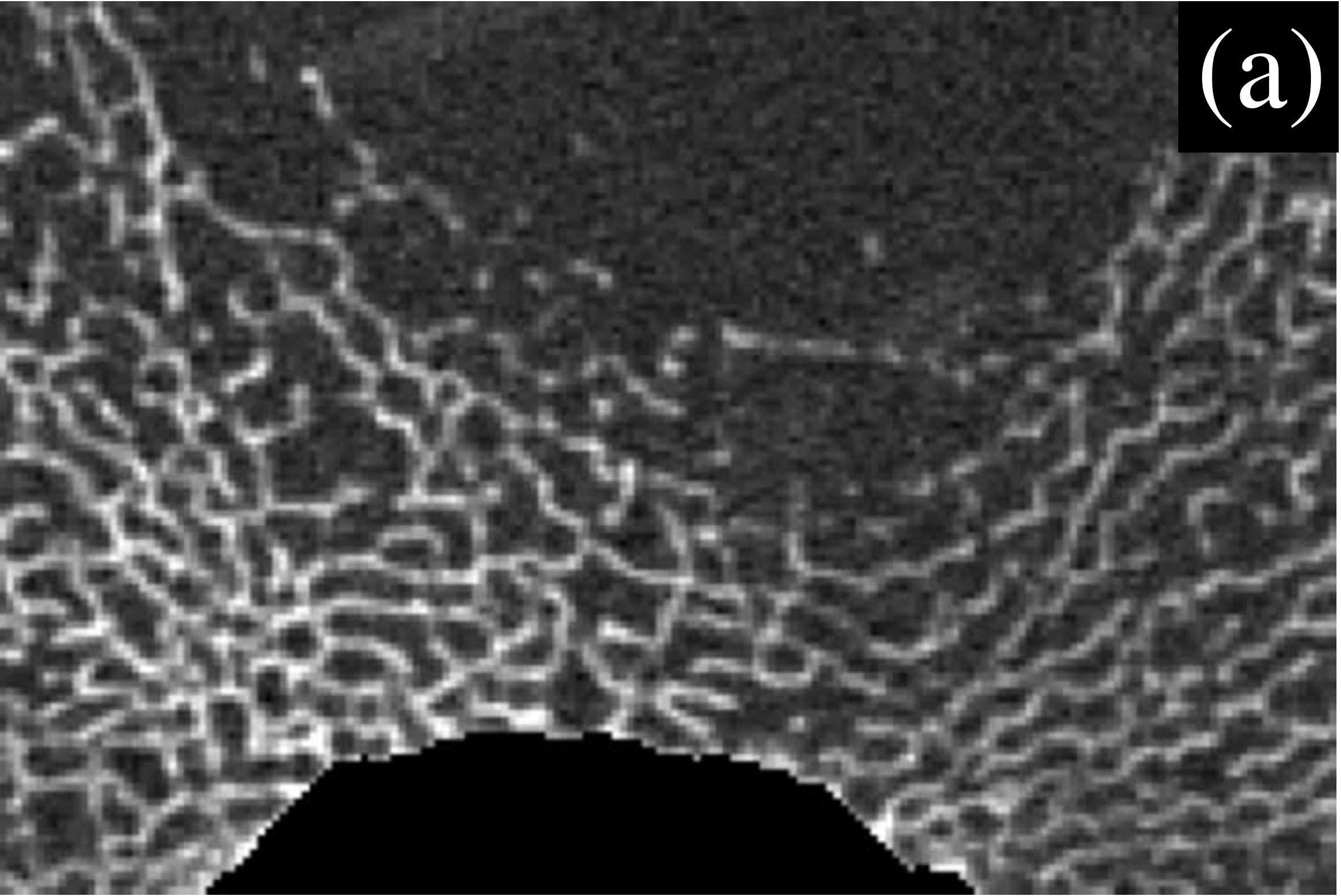} 
\includegraphics[height=.3\textwidth,angle=-90]{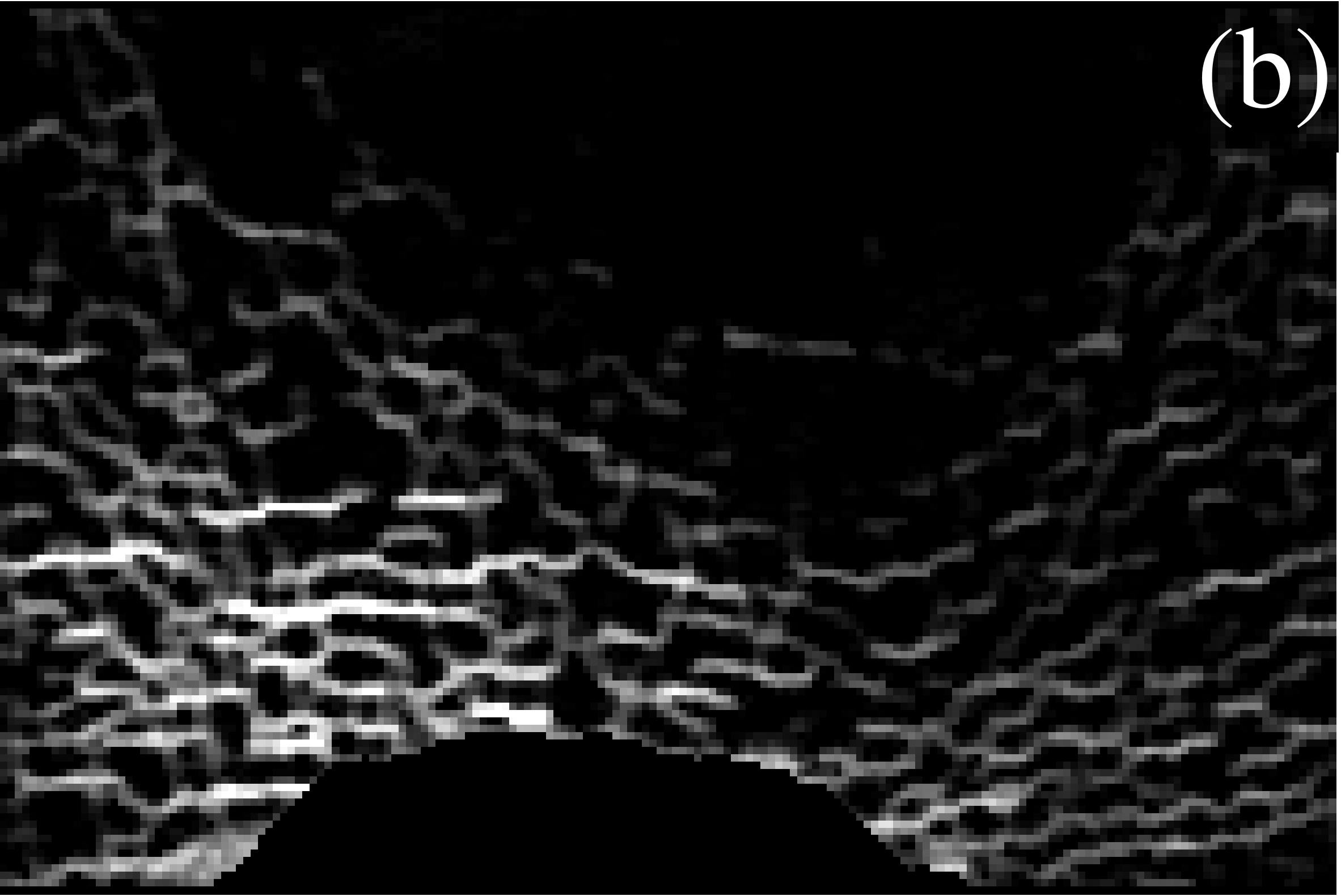} 
\includegraphics[height=.3\textwidth,angle=-90]{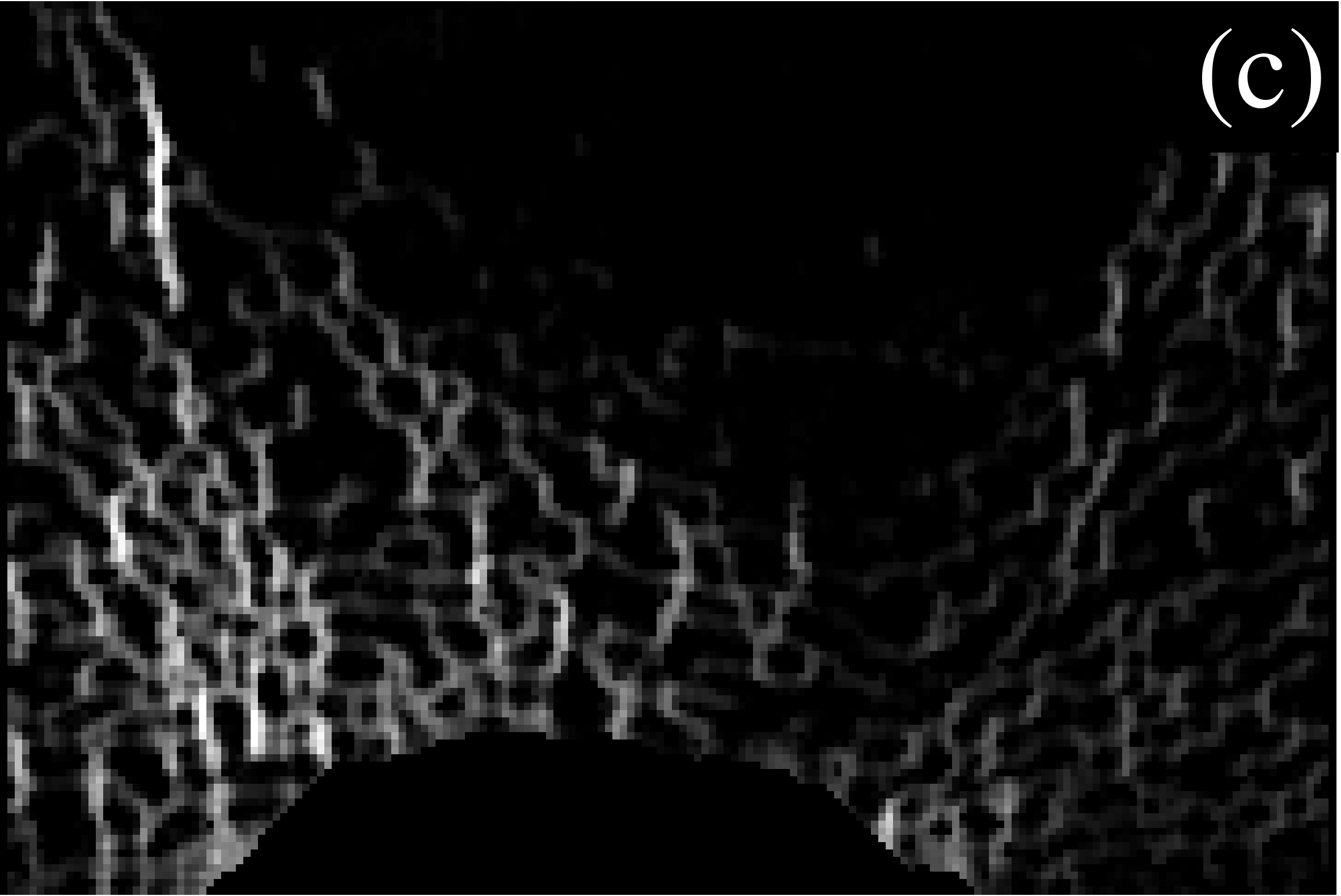} \\ \vspace{3pt}
\includegraphics[height=.3\textwidth,angle=-90]{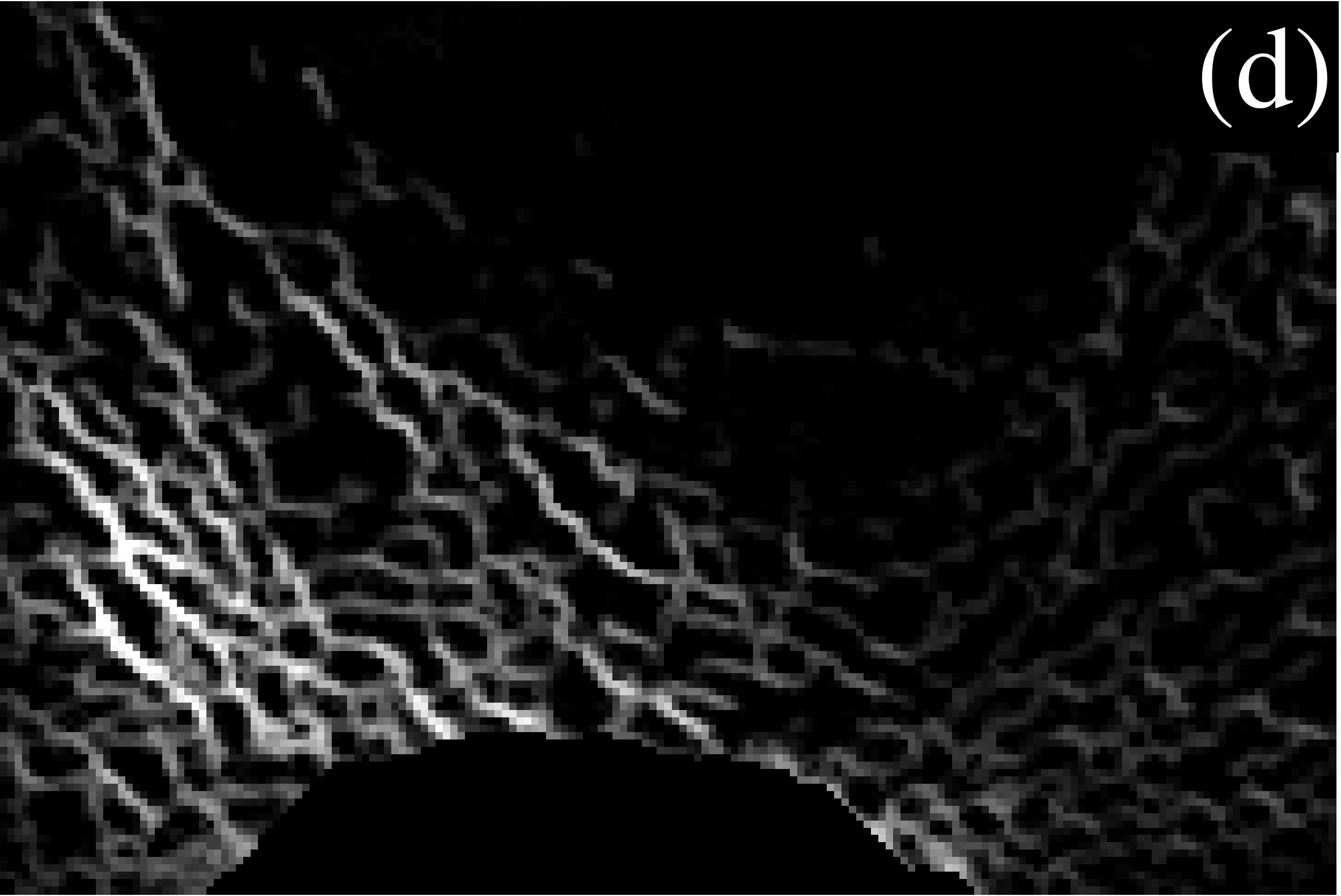} 
\includegraphics[height=.3\textwidth,angle=-90]{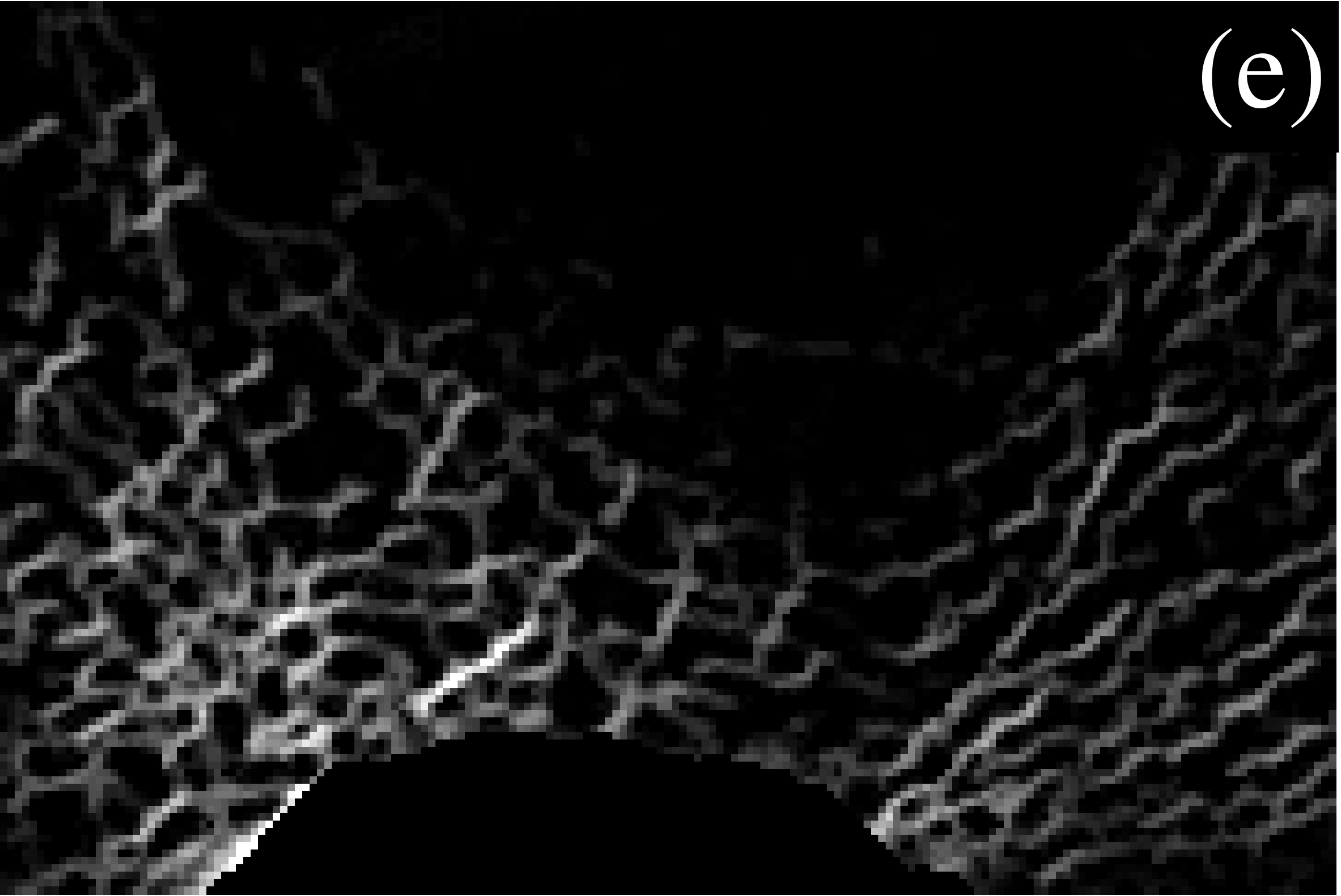} 
\includegraphics[height=.3\textwidth,angle=-90]{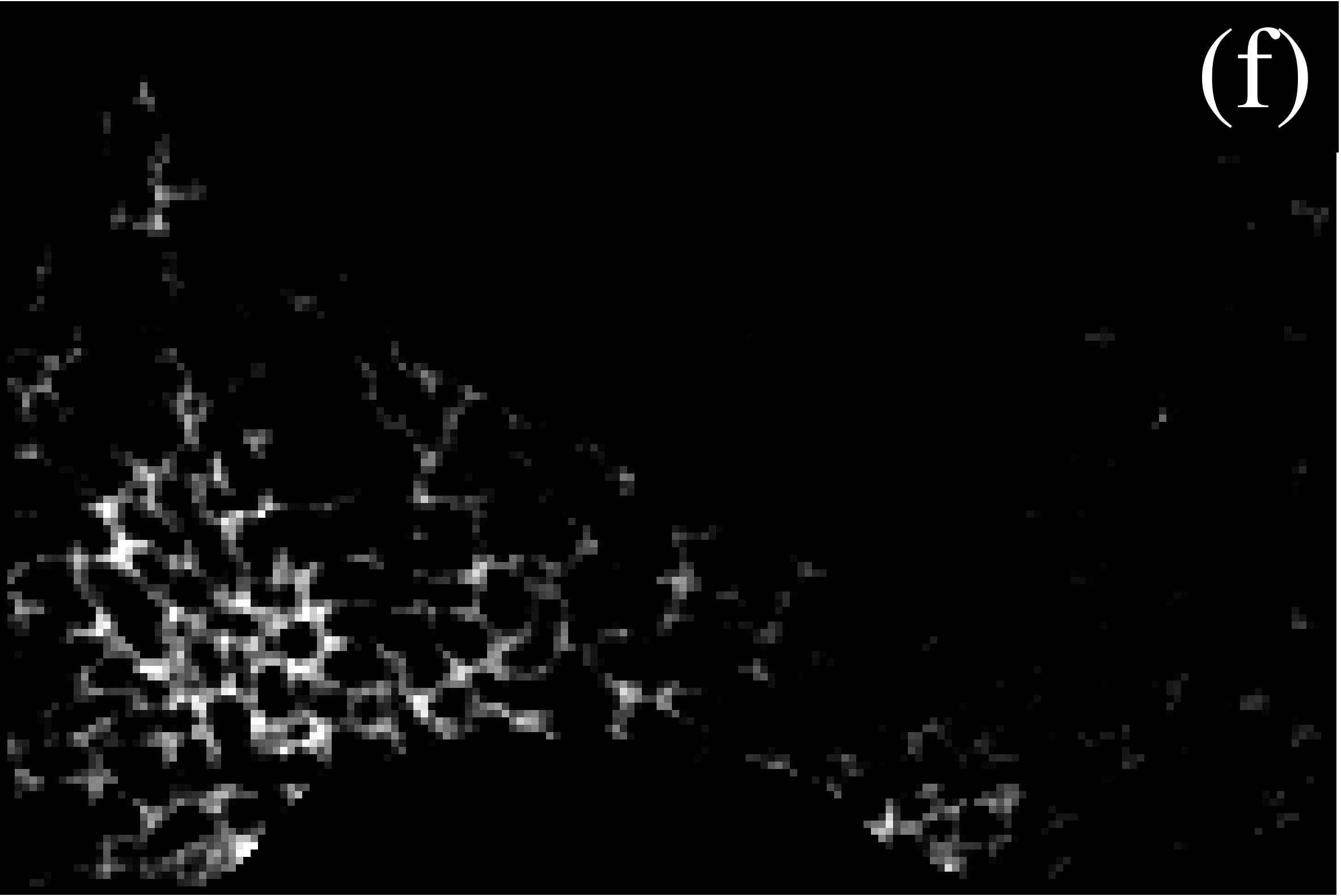} 
\caption{
Node strength algorithm illustrated on a region of trabecular bone
taken from the full section shown in Fig.~\ref{fig:tibia}.
(a) original image; (b) horizontal strands; (c) vertical strands; 
(d) diagonal strands (``northwest - southeast'');
(e) diagonal strands (``northeast - southwest''); (f) node strength.
}
\label{fig:patch}
\end{figure*}

Finally, we calculate the node strength $\sigma_{ij}$ at each pixel. 
There are $k=16$ possible ways of choosing
3 out of the 8 directions in such a way that each pair of chosen directions
makes an angle of at least 90 degrees.
For example, E, N, and SW comprise an allowable choice,
but E, N, and NE do not.
At each pixel, for each allowable choice of three directions, we calculate the minimum of the three strand strengths in these directions, e.g., 
$\hat{S}_{ij}^k = \min \{S^E_{ij},\ S^N_{ij},\ S^{SW}_{ij}\}$.
The \emph{node strength} $\sigma_{ij}$ is the maximum 
of the 16 minima $\hat{S}_{ij}^k$ subtracted by
a minimum strength constant $S_0$,
\begin{equation}
\label{eq_ndstr}
\sigma_{ij} = \max_{k=1\ldots 16} \big\{\hat{S}_{ij}^k\big\} - S_0.
\end{equation}
Pixels with a positive node strength are called \emph{nodes}.
The mean node strength over the region of interest (ROI) is called the node strength of the region, 
\begin{equation}
\label{eq_meanndstr}
\text{NdStr} = \frac{1}{\text{ROI}}\sum_{ij \in \text{ROI}} \sigma_{ij}.
\end{equation}

The purpose of subtracting the minimum strength constant $S_0$ is to
allow us to ignore short  ``strands'', which are often just transverse sections
of trabeculae with an apparent width of more than one pixel.
The trabeculae in our images have an apparent width of approximately
two pixels, or 0.4\,mm,
which is higher than the true average trabecular width, due to the
1\,mm thickness of the CT image and the 0.2\,mm pixel size.
Thus, we wish to ignore strands of bone that are only two pixels
long. The strength of such a strand depends on the pQCT values of the
two pixels. 
Following Eq.~(\ref{E:strandstr}), the strength $s$ of 
such a strand with CT value $b_0=b_1=b$ is $s=b$. 
Based on examination of the resulting node strength
images (such as Fig.~\ref{fig:patch} (f)), we find ad hoc 
a minimum strength constant of 225 (corresponding to a
CT value of 500, Eq.~(\ref{E:threshold}), corresponding to a BMD value of 331\,mg/cm$^3$,
which is higher than the mean CT value for trabecular bone). 
So strands of lower strength than this will be ignored by the algorithm.

We note that our complete algorithm involves two thresholding steps: one
at the beginning, when pQCT values are converted to CT values of bone;
and one at the end, when the minimum strength constant is subtracted.
These are not equivalent to one thresholding step with a higher threshold.
The purpose of the first threshold filtration is to ignore marrow.
The purpose of the second threshold filtration is to ignore
short strands, and it can be seen as an alternative to skeletonisation.

Numerical experiments have shown that relative (cross-subject) values of
NdStr are stable with respect to the choice of the bone threshold value
($a^\textrm{threshold}$) and minimum strength constant. Specifically, if
either of these constants is varied by $\pm 10\%$ from the values used in
this article, and NdStr is recomputed for all subjects, the resulting
NdStr values are very strongly linearly correlated with the original NdStr
values (Pearson correlation coefficient $r > 0.995$ in all cases). The
absolute values of NdStr depend on the choice of these constants:
increasing the bone threshold by $10\%$ leads to a $40\%$ mean decrease in
NdStr (percent decrease averaged over all subjects), while increasing the
minimum strength constant by $10\%$ leads to an average decrease in NdStr
of $11\%$. Nevertheless, and as already mentioned, the relative 
(cross-subject) variation of NdStr remains constant.

A key property of NdStr
is that it depends on both
the geometry of the trabecular network and the CT values of the trabeculae.
NdStr depends linearly on thresholded CT values, \emph{if} the
CT values of all pixels are varied by the same factor. 
It follows that images with the same geometry but different BMD will have different NdStr values.
On the other hand, two images with the same BMD but different geometry can have different NdStr values.
This is apparent from the description of the method, and confirmed by
Figs.~\ref{fig4} and \ref{fig5}, as discussed below, and also by the
results discussed in Section \ref{sectResults}.

\medskip

\subsection{Illustration}

\begin{figure*}[htbp] 
\centering \includegraphics[width=.7\textwidth]{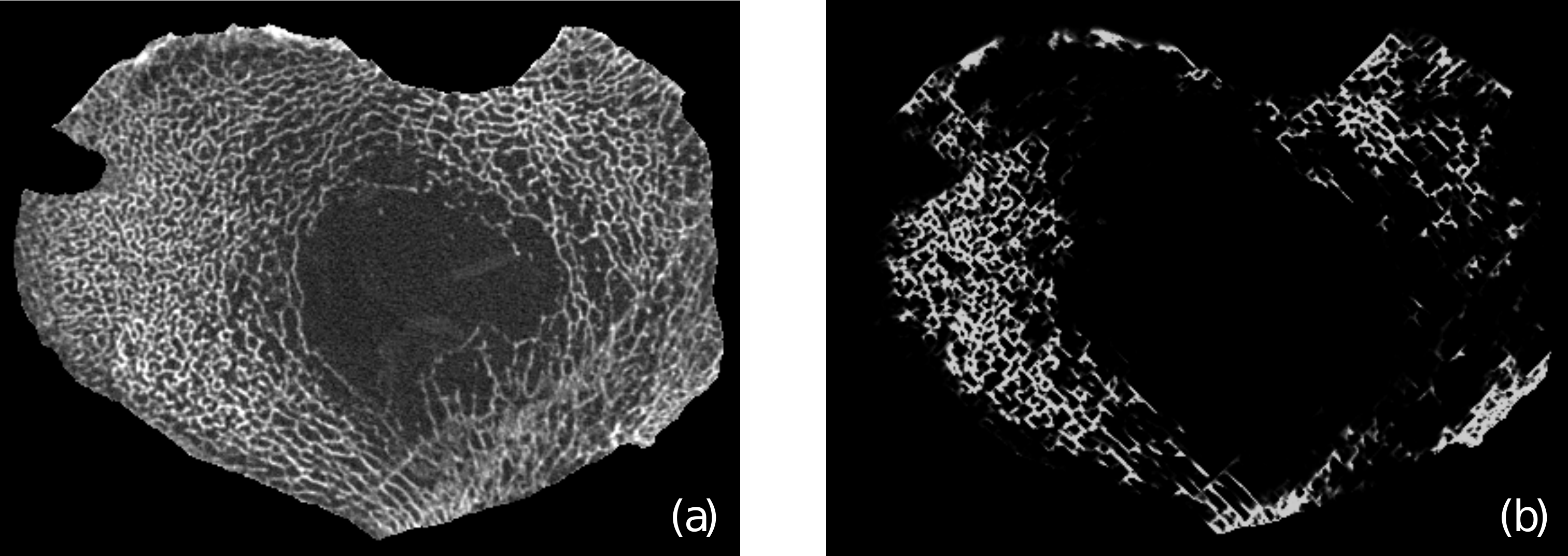} 
\caption{Trabecular micro-architecture of a section with BMD
107\,mg/cm$^3$. (a) original image; (b) the node strength of each pixel.
The mean node strength is 71.2.
}
\label{fig4}
\end{figure*}

To illustrate the analytical method, 
we now present the results in visual form for 
an enlarged region near the bottom (posterior)
of the slice in Fig.~\ref{fig:tibia}.
The enlarged region of the original image is shown in 
Fig.~\ref{fig:patch}(a).
Parts (b) to (e) of Fig.~\ref{fig:patch} show
directional strand strengths, and part (f) shows the final node strength plot.
Each directional strand strength plot shows the sum of two strand strengths
at every pixel,
in opposite directions: east/west;  north/south; northwest/southeast;
and northeast/southwest.
In each of the directional strand strength plots, the strands in the
given direction are shown with the highest intensity, but most of the trabeculae are
still visible, even if faintly.
In contrast, in the node strength plot (part (f)), most of the trabeculae are invisible.
This is because of the subtraction of the minimum strength constant.
In this example, there are almost no nodes in the right half of the image.
This correctly describes the micro-architecture of the original image in that region,
which contains many trabeculae but few that cross each other to make a
lattice-like micro-architecture.
The left half of the image contains many nodes. Notice that, in the node strength plot,
the nodes seem to be thicker than in the original image. 
This is because the trabeculae in the original image are actually slightly thicker than they
appear, the outer pixels being dimmer (i.e.~lower CT values) 
and thus not easily registered by the eye.
Since the outer pixels near the apparent nodes in the original image are almost
as well-connected as pixels in the centres of the nodes,
they have large node strengths, and are very visible in the node strength plot.

\begin{figure*}[htbp] 
\vspace{0.2cm}
\centering \includegraphics[width=.7\textwidth]{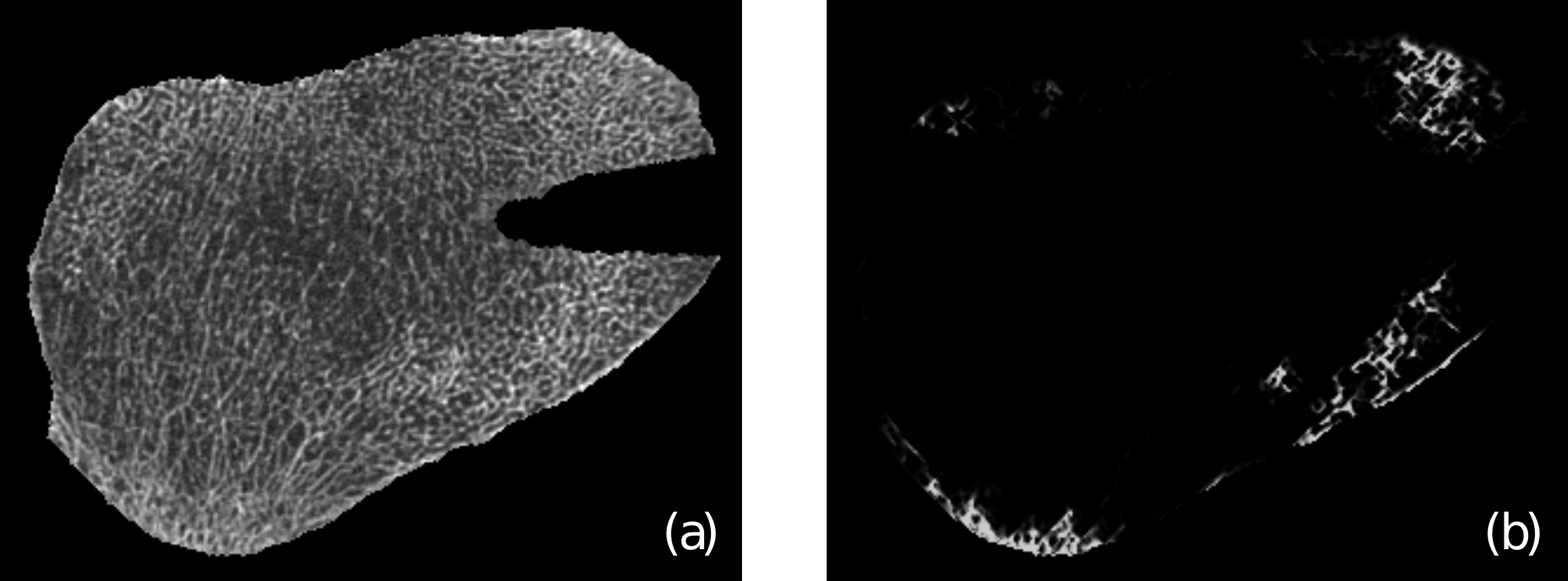} 
\caption{
Trabecular micro-architecture of a section with BMD
is 94\,mg/cm$^3$. (a) original image; (b) the node strength of each pixel.
The mean node strength is 42.2.}
\label{fig5}
\end{figure*}

The two specimens depicted in Figs.~\ref{fig4} and \ref{fig5} 
have comparable bone mineral densities,
but their trabecular micro-architecture are visibly different.
In Fig.~\ref{fig4}, the specimen has a
trabecular BMD of 107\,mg/cm$^3$, which is near the median value 
(97\,mg/cm$^3$) of the specimens in this study.
On the left is the original image, and on the right is the node strength plot.
Notice that there are a lot of nodes in most of the outer areas,
with the notable exception of a region near the bottom left.
The mean node strength is 71.2.
%

The specimen in Fig.~\ref{fig5}
has a trabecular BMD of
94\, mg/cm$^3$, 
which is only 12\% lower than that of the specimen shown in Fig.~\ref{fig4}, but
it has substantially fewer nodes.
The mean node strength is only 42.2, 
which is 40\% lower than that for the specimen shown in Fig.~\ref{fig4}.
This reflects the lack of a strong lattice-like micro-architecture in the original image.


\section{Results}\label{sectResults}

\begin{figure}[h] 
\centering \includegraphics[width=\columnwidth]{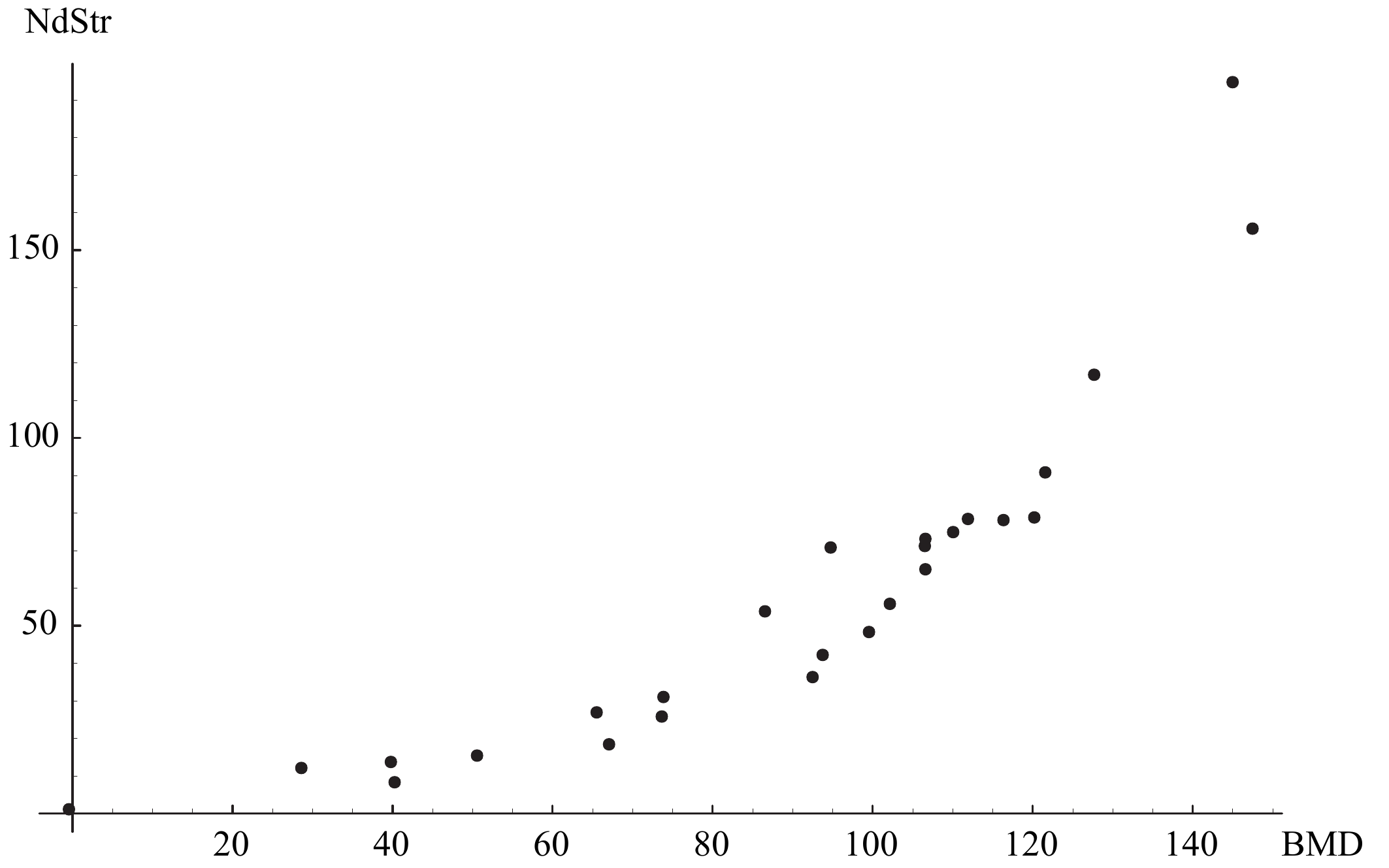} 
\caption{
Scatter plot of mean node strength vs.~trabecular bone mineral density 
(mg/cm$^3$).
}
\label{fig:nodesbmd}
\end{figure}

The long range node strength of each pixel of each image 
was computed as described in the previous section,
and a mean node strength, NdStr, was calculated for each image.
We also computed the bone mineral density (BMD)
from the pQCT slices, as explained in Section \ref{sectAnalysis}.
A scatter plot showing NdStr versus BMD for
all 26 specimens is shown in Fig.~\ref{fig:nodesbmd}. 
There is a strong positive correlation, which we
quantified in three ways. Firstly,
Pearson's correlation coefficient is $r=0.87$, indicating a very strong linear correlation.
Secondly, since the scatter plot clearly suggests a nonlinear relationship,
we fitted an exponential curve to the data and
then found
Pearson's correlation coefficient to be $r=0.97$.
Thirdly, the Spearman rank correlation is $\rho=0.98$,
indicating a very strong correlation.
The Spearman rank correlation coefficient is a robust nonparametric correlation measure that is 
appropriate when little is known about the distributions and nature of the correlation
between the variables. \cite{Press1992}

We also compared NdStr with 
the node-terminus ratio, Nd/Tm, in the topological
node-strut analysis introduced by Garrahan et al. \cite{Garrahan1986}
The two measures are similar in philosophy, because they both quantify the nodes
in the trabecular network.
However, the definition of Nd/Tm is highly localised:
after the skeletonisation process has eroded the trabecular network to a thickness
of one pixel, each pixel is classified as node, strut, or terminus depending
on its $3\times 3$ grid of nearest neighbours (including the original pixel).
For the present study, the images used to compute Nd/Tm have a  pixel size of
$10\, \mu$m, so the classification is made on the basis of a 
30\,$\mu$m $\times$ 30\,$\mu$m region.
In contrast, node strength is semi-global, 
taking into account longer strands to a degree controlled by the transmission
strength constant. In the present study, with this constant set to $0.95$,
the method has a characteristic length of 4~mm, in the sense described in 
Section \ref{sectAnalysis}.

The node-terminus ratio, Nd/Tm, was calculated using histomorphometry performed
on bone biopsies as described in Section \ref{sectAnalysis}.
Recall that these biopsies were from the same regions of the same donors
as the pQCT slices from which NdStr has been computed.
Pearson's correlation coefficient for the relationship between NdStr and Nd/Tm is $r=0.62$, and
the Spearman rank correlation coefficient is $\rho = 0.65$.
We also measured the correlation between Nd/Tm and
trabecular BMD:
Pearson's correlation coefficient is $r=0.64$, and
the Spearman rank correlation coefficient is $\rho = 0.61$.
Using either measure of correlation, mean node strength is more strongly correlated with 
trabecular BMD than the node-terminus ratio is correlated with either
of these variables.

%

\section{Discussion}
\label{sectDiscussion}

We have introduced a new morphometric measure for characterising the micro-architecture
of trabecular bone,
\emph{long range node strength}, 
which measures the degree to which a pixel
in a 2-dimensional bone image has long-range connectivity in three or more
directions, each at least 90 degrees from the others.
In addition, we have calculated the mean node strength, NdStr, 
for each of the 26 bone samples considered in the study.
We have found that NdStr has
a strong positive correlation with trabecular BMD ($r=0.97$, after exponential
transformation).
Furthermore, we have ascertained a strong correlation ($r= 0.62$) between NdStr 
and the established histomorphometric measure, node-terminus ratio (Nd/Tm).
Moreover, qualitative comparison of images with similar trabecular BMD but different
mean NdStr (see Figs.~\ref{fig4} and \ref{fig5}) suggest that NdStr 
successfully quantifies how ``lattice-like'' the micro-architecture is.


Further studies, including either clinical data or measured bone strength,
are needed in order to determine the utility of this measure relative to
other existing morphometric measures.
Such studies could also determine the most useful choice
of the parameters that appear in the algorithm:
the transmission constant, minimum strength constant, and bending coefficients.
For example, these constants could be chosen to maximise
correlations with bone strength.
In the future, the sensitivity of the method
on these parameters and on different CT settings
(e.g., CT pixel size, slice thickness, mean CT value of bone)
should be systematically investigated.
Moreover, a future study on synthetic data could be used to verify
that NdStr measures structure and not just bone volume
or mass, and specifically that it finds nodes.
Also different skeletal sites
should be analysed with our approach
in order to test its potential to describe structural
differences.

We would like to note that our algorithm for computing node strength
is dependent on pixel size, and is thus unsuitable for absolute
comparisons between studies.
However, this limitation is not unique to the node strength measure. For example,
Guggenbuhl \emph{et al.} have showed using CT images with
different thickness (1 mm, 3 mm, 5 mm, and 8 mm) that the outcome of texture analysis
depends substantially on the slice thickness\cite{Guggenbuhl2008}.
In the present study we have used pQCT equipment with an in-plane  pixel size
of 200 $\mu$m $\times$ 200 $\mu$m and a slice thickness of 1 mm. 
We have not conducted a formal investigation into the influence of slice
thickness on the long range node strength, but it is fair to assume that
the long range node strength is similarly affected by the slice thickness.
If further studies were to confirm the practical value of the measure,
an implementation could be developed to produce node strengths that are
broadly comparable between images with different  pixel size.
However, the technological development of high-resolution pQCT equipment
is progressing very quickly, and already high-resolution pQCT scanners exist
that can image a tibia at an isotropic  pixel size of approximately 100 $\mu$m,
which is comparable to the trabecular thickness in the human proximal
tibia\cite{ding2000}, thus making the slice thickness less of an issue.


Some previous studies have investigated the trabecular bone micro-architecture
using texture analysis applied on X-ray images of
bone\cite{Geraets1990,Chappard2005b,Chappard2006,Huber2009,Korstjens1996}.
However, these techniques are not comparable to the method presented in the present
study as they are based on projections of a 3-dimensional trabecular network on a
2-dimensional plane, whereas our method is applied to 2-dimensional sections obtained
through the 3-dimensional trabecular network.
Nevertheless, Ranjanomennahary \emph{et al.} very recently showed some significant correlation
did exist between radiograph based texture analysis and $\mu$CT based unbiased 3-dimensional
measures of trabecular micro-architecture\cite{Ranjanomennahary2011}.
Apostol \emph{et al.} compared 3-dimensional measures of micro-architecture based on 3-dimensional
synchrotron radiation CT with more than 350 texture parameters obtained from simulated radiographs
that were created from the synchrotron radiation CT data sets\cite{Apostol2006}.
They found using multiple regression analysis that a combination of a subset of texture
analysis parameters correlated to the 3-dimensional measures of micro-architecture. However, they
had to use at least three texture analysis parameters in the correlation with each of
the 3-dimensional measures of micro-architecture. 
Cortet \emph{et al.} investigated CT images of the distal radius using texture
analysis\cite{Cortet2000,Cortet1999}.
The CT image used by Cortet \emph{et al.} used similar  pixel size to those used in the present
study.
They analysed the CT images using the traditional node-strut analysis and texture analysis
including the grey level run length method\cite{Chu1990,Galloway1975}.
As illustrated in the present work there is a moderate correlation ($r = 0.64$) between
the node-strut analysis and the NdStr measure, which we believe illustrates that the
NdStr measure captures somewhat different information from that obtained with the
node-strut analysis.
Grey level run length is based on runs that have exactly the same grey level, and is
therefore very sensitive to the choice of discretisation of the grey levels.
In contrast, the NdStr measure treats the grey level as a continuous parameter, and
thus it is able to detect runs with variable grey levels, and is consequently less
sensitive to discretisation choices.

Another advantage of the method is that the proposed method utilises all of the 
CT value information in the pQCT images, as no binarization of the images is performed.

In the present study we have not applied the developed method to the histological
sections directly. The reason for this is that the histological sections only
cover a limited area of interest and thereby a limited trabecular length, which
renders the NdStr less meaningful. However, this is a limitation of the histological
examination procedure where it is only feasible to investigate a smaller sample (biopsy)
of a larger structure (the proximal tibia) and not a limitation of the proposed method.
In addition, the histological images are segmented into bone and marrow
by their nature and it is thus not possible to assign an intensity value
to pixels, analogous to a CT value, which is needed by the algorithm in its present form. 
However, it is possible to apply the concept of long range node-strut
analysis to such binary images but this is outside the scope of the 
present investigation.

In the present study the long range node-strut analysis was applied to
pQCT images of the proximal tibia. However, we would like to stress that
the method is not limited to this skeletal site and thus can be applied
to 2-dimensional CT images obtained from any skeletal site like e.g. the
vertebral body or the calcaneous.

Finally, we note that the general method of long range node-strut analysis 
provides more than just the mean node strength. In the present study, we have 
focused on mean node strength for simplicity,
but the intermediate measures of \emph{directional strand strength}, 
used in the computation of node strength,
may be useful in themselves as a measure of directional strength.
In the present study the long range node-strut analysis has been applied
to 2-dimensional pQCT images obtained in the horizontal plane.
However, the trabecular micro-architecture of the proximal tibia is mostly
isotropic in the horizontal plane, whereas the micro-architecture in vertical
direction is highly anisotropic compared with the horizontal plane\cite{ding2002}.
As $\mu$CT scanners become more and more prevalent and as the
pixel size and imaging capacity of pQCT scanners steadily improve it will
be an important future task to generalise the long range node-strut
analysis into three dimensions and to apply the technique to 3-dimensional
data sets obtained from such equipment. In particular, the directional strand
strength could be used to investigate anisotropic differences of the
trabecular micro-architecture of such 3-dimensional data sets.


\section*{Acknowledgements}

The data acquisition parts of this project were made possible by grants from the Microgravity Application Program/Biotechnology from the Manned Spaceflight Program of the European Space Agency (ESA) (ESA project \#14592, MAP AO-99-030). The authors would like to thank Professor G.~Bogusch and Professor R.~Graf, Center for Anatomy, Charit\'{e} Berlin, Germany for kindly providing the bone specimens. Dr. Wolfgang Gowin, formerly at Campus Benjamin Franklin, Charit\'{e} Berlin, Germany, is gratefully acknowledged for preparing the bone specimens and harvesting the bone biopsies. Erika May and Martina Kratzsch, Campus Benjamin Franklin, Charit\'{e} Berlin and Inger Vang Magnussen, University of Aarhus, Denmark are acknowledged for their excellent technical assistance scanning the CT images and preparing the histological sections.



%

\end{document}